\documentclass{pasj00}
\draft

\begin{document}
\SetRunningHead{M. Nakano et al.}{Emission-line Stars in the W5E region}
\Received{2007/11/19}
\Accepted{2008/2/18}

\title{Clustering of Emission-line Stars in the W5E HII region}

\author{Makoto \textsc{Nakano} %
}
\affil{Faculty of Education and Welfare Science, Oita University,
Oita 870-1192, Japan}
\email{mnakano@cc.oita-u.ac.jp}
\author{Koji \textsc{Sugitani}}
\affil{Graduate School of Natural Sciences, Nagoya City University,
Mizuho-ku, Nagoya 467-8501, Japan}
\author{Takahiro \textsc{Niwa}, Yoichi \textsc{Itoh}}
\affil{Graduate School of Science and Technology, Kobe University, Kobe 657-8501, Japan}
\and
\author{Makoto \textsc{Watanabe}}
\affil{Subaru Telescope, National Astronomical Observatory of Japan, 650 North A'ohoku Place, Hilo, HI 96720, USA}

%


\maketitle

\begin{abstract}
We have made a new survey of emission-line stars in the W5E HII region to  investigate the population of 
PMS stars near the OB stars by using the Wide Field Grism Spectrograph 2 (WFGS2).
A total of 139 H$\alpha$ emission stars were detected and their g'i'-photometry was performed.
The spatial distribution of them shows three aggregates, i.e., two aggregates near the bright-rimmed clouds at the edge of 
W5E  HII region (BRC 13 and BRC 14) and one near the exciting O7V star.
The age and mass of each H$\alpha$ star were estimated from the extinction corrected color-magnitude diagram 
and theoretical evolutionary tracks.
We found, for the first time in this region,  that the young stars near the exciting star are systematically older (4 Myr) than
those near the edge of the HII region (1 Myr).
This result supports that the formation of stars proceed sequentially from the center of HII region to the eastern bright rim.
We further suggest a possibility  that the birth of low mass stars near the exciting star of HII region
precede the production of massive OB stars in the pre-existing molecular cloud.
\end{abstract}

\section{Introduction}

On the basis of the field star initial mass function (IMF) \citep{mis79}, one would expect an OB-type 
star to form along with a few hundred low mass objects.
Thus, low mass members of OB associations are an important key to understand the main formation process of
stars in our Galaxy, but  it is not easy to find them observationally \citep{bri07}.
As they are too faint to observe and widely distributed in space, it is inevitable to avoid the contamination by field stars. 
The systematic search for low-mass members of OB associations demonstrated that some external 
agents are needed for the coordinated formation of stars \citep{pre02} or the termination of the low-mass 
star formation \citep{dol02}.
So far a large scale triggering of stellar birth by external pressure from HII regions,
stellar winds, and supernovae have been proposed for the formation of subgroups of
OB associations  (e.g.,\cite{pre07}).  
On the other hand, a small scale triggering also seems operating in several other cases.
Bright-rimmed clouds (BRCs) are typical sites of such a small scale triggering
by high pressure squeezing of the pre-existing clumps at the edge of HII regions.
Many observational works indicated that they are active sites of low-to-intermediate mass star formation.
It is difficult to estimate how a small scale triggering affects to the IMF,  but
in some cases the triggering process may work to IMF locally (e.g., \cite{get07}).

The Cas OB6 association extends over several degrees and is made up of a large number of
active star forming regions spread along the Perseus arm of the Galaxy  and
is located at a distance of about 2.3 kpc (\cite{gar92}; \cite{mas95}).
The chain of HII regions W3/W4/W5 extends over 150 pc from west to east.  
\citet{car00} found  evidence of many embedded star forming sites in the W3/W4/W5 regions 
using the IRAS Point Source Catalog.  They studied young clusters associated with the giant 
molecular cloud, and discussed the triggered star formation hypothesis.
W5  consists of two main components of the ionized gas, the large western part (W5W) 
and the circular eastern part (W5E) of HII region. 
O7V star HD18326 (BD +59 578) is located near the center of W5E and the primary 
source of ionizing flux. 
W5E  shows  rather simple round structure with 40' in diameter, which corresponds to 27 pc,  
but  has several bright rims along its eastern and northern perimeters.

\citet{kam03} discussed the clustering lengths of high-to-intermediate mass young stellar 
objects selected from IRAS sources and the associated timescales, and suggested
the formation of stars in the W5 region was triggered at the edge of the expanding HII region. 
Also, an optical and infrared study \citep{deh97} of the prominent bright rim  
BRC 14 or SFO14 \citep{sfo91}, which is at the eastern rim of W5E, revealed an active star formation site.  
They studied the massive pre-main sequence (PMS) object AFGL 4029-IRS1 (IRAS 02575+6017),
which is deeply embedded (Av=25--30 mag) and luminous ($>$ 10$^4$ L$_\odot$),
illuminating a reflection nebula as a part of a rich star cluster. 
They suggested about 10\% of the mass of the parental cloud having been used to
form B stars.
A deeper near-infrared survey of faint YSO candidates of this cluster  was conducted by 
\citet{mat06}.  
They examined three indicators of stellar youth in three selected areas from 
outside the rim to the molecular cloud, and showed 
all of them support the triggered formation of stars.

NIR survey is a powerful method to search for embedded PMS cluster inside molecular clouds  (e.g.,\cite{lad03}),
but biased toward PMS stars with NIR excess.
On the other hand, H$\alpha$ survey is biased toward those with strong H$\alpha$ emission  and low
extinction.
However, H$\alpha$ emission is readily detectable on grism spectrograms at moderate
equivalent width. Thus H$\alpha$ survey remains an efficient and reliable way to identify PMS objects.

In this paper, we present the result of a wide field survey of emission-line stars in the W5E region.
H$\alpha$ emission star is a good candidate of the PMS star, 
and a convenient probe for investigation of the star formation history.
If the exciting star of HII region was formed together with low mass stars, we could
find the young low mass members in the HII region.
Young members, often found in cluster,  also provide a fossil record to understand  
the timescale of star formation in giant molecular clouds.   
Therefore we have  investigated the population of PMS stars in large area including the inside 
of HII region and very near the OB stars, and approach the sequence of star formation of this
region.

\newpage

\section{Observations}

The H$\alpha$ survey and the optical photometry of the region were conducted in 2006 September
22 and 24. 
The sky condition during the run was mostly photometric, but the second night a few cirrus cloud
was observed.
We used the Wide Field Grism Spectrograph 2 (WFGS2; Uehara et al. 2004) mounted on 
the University of Hawaii 2.2-m telescope.  The detector was a Tektronix   2048$\times$2048  CCD, 
which gave a pixel scale of 0.34" pixel$^{-1}$ and an instrumental field of view of 11.5' $\times$ 11.5' at the focus f/10 of the telescope.   
The W5E region including two conspicuous bright rims (BRC 13 or SFO13 and BRC 14) and the exciting source 
was observed at four field positions both in the grism and the direct imaging mode of WFGS2.
Table 1 shows the coordinates of center of the observed fields.

\begin{table*}
\begin{center}
\caption{Observed fields}
\begin{tabular}{lrrc}
\hline
\hline
  Field & RA(2000) & Dec(2000) & Number of H$\alpha$ stars  \\
  \hline
 W5E-a &  3:01:09.9 & +60:39:28.0 & 16 \\
 W5E-b & 3:01:38.2  & +60:29:00.3 & 45 \\
 W5E-c &  3:00:16.9 & +60:31:57.5 & 52 \\
 W5E-d &  2:59:23.0 & +60:35:24.0 & 40 \\
\hline
\end{tabular}
\end{center}
\end{table*}

In the grism mode, we obtained three 180 sec  exposures for W5E-a field, and three 300 sec  exposures for each of W5E-b,c,d 
fields for the slit-less spectroscopy with a wide-H$\alpha$ filter (FWHM=50nm) and a 300 line mm$^{-1}$ grism.
Direct images were also taken with  g' and i'-band filters.
Three 40 sec exposures for  g'-band, and  three 30 sec exposures for i'-band were made.
We used twilight images as flat fielding.
Roughly 3--5 standards for the Sloan Digital Sky  Survey (SDSS) system were observed 
each night nearly at the same airmass  ( difference of $<$ 0.1) as the target. 

The data were reduced by using the IRAF software package.
At first, we have picked up stars in the short exposure H$\alpha$ images by the DAOFIND package,
and extracted the spectra of them from grism images.
One dimensional spectra of stars were investigated for the presence of H$\alpha$ emission line by eye.
After three times of inspection, we have found 16, 45, 52, and 40 H$\alpha$ stars in W5E-a, W5E-b, W5E-c, 
and W5E-d, respectively.
We used the SPLOT package to measure the equivalent width of H$\alpha$ line in the spectra.
Their coordinates were determined from the pixel coordinates provided by the IRAF routines and  
the values of RA, Dec of a network of nearly 10  reference stars, whose positions were referred to 
USNO-B catalog stars in the field.

All the photometric measurements were made with the DAOPHOT package for aperture photometry.
The FWHM of the image is 2.7--3.0 pixel.
In each field, $\sim$800--1800, and $\sim$2300--3800 stars were measured above 5$\sigma$ of the sky level 
with an accuracy of less than 0.1 mag in  g'- and i'-band, respectively.
To correct the photometric quality by the variation of sky condition in the second night, we used the stars in 
the overlapped region between the fields to adjust the photometric zero points to the first night ( $<$ 0.1 mag ).
We then calibrated the stars to SDSS system via standard from \citet{smi02}.
Ninety-one of H$\alpha$ stars were measured in two color bands.
We used some stars in BRC 14 observed 
by \citet{deh97} to confirm our photometric quality. 
The limiting magnitude at  10 $\sigma$ level  is slightly different depending on the seeing condition, i.e.,
21.9--22.1 mag in g'-band, and 21.2--21.6 mag in i'-band. 

\section{Results}

\subsection{Spatial Distribution of Emission-line Stars}

As 14 stars are overlapped in the field, we detected a total of 139 H$\alpha$ emission stars.
Figure 1 illustrates the spatial distribution of the detected H$\alpha$ emission stars.
Figure 2 shows isodensity contours of the surface density  of the H$\alpha$ emission-line stars
with the kernel method \citep{gom93}. 
We adopted a Gaussian shape for the density distribution of the kernel 
with a smoothing parameter h=1.2'.
Three groups associated with BRC 13, BRC 14, and the central exciting star  HD18326 are
found.  We denote them as the group A, B, and C for the following discussion.
The adopted boundary of each group is shown in figure 2.
Table 2 shows the number of H$\alpha$ stars in each group.

\begin{table*}
\begin{center}
\caption{Detected H$\alpha$ stars}
\begin{tabular}{cccl}
\hline
\hline
  Group & H$\alpha$ stars & EW$<$10 \AA & Remarks \\
 \hline
  A &  16 & 2 & BRC 13(SFO13) \\
  B &  56 & 9 & BRC 14(SFO14) \\
  C &  67 & 11 & around HD18326 \\
\hline
\end{tabular}
\end{center}
\end{table*}

\citet{ogu02} searched H$\alpha$ stars and HH objects in and around 26  bright-rimmed clouds
listed mainly in the catalog of \citet{sfo91} and \citet{so94}.  
They detected as many as 460 H$\alpha$ emission stars, and among them
24 and 47 H$\alpha$ stars are near BRC 13 and 14, respectively.
About half of them are confirmed by our robust identification.
120 (86 \%) of our H$\alpha$ stars are identified with the 2MASS sources, and two thirds of them
has good photometric quality in J, H, K$_{s}$-bands.

\begin{figure}
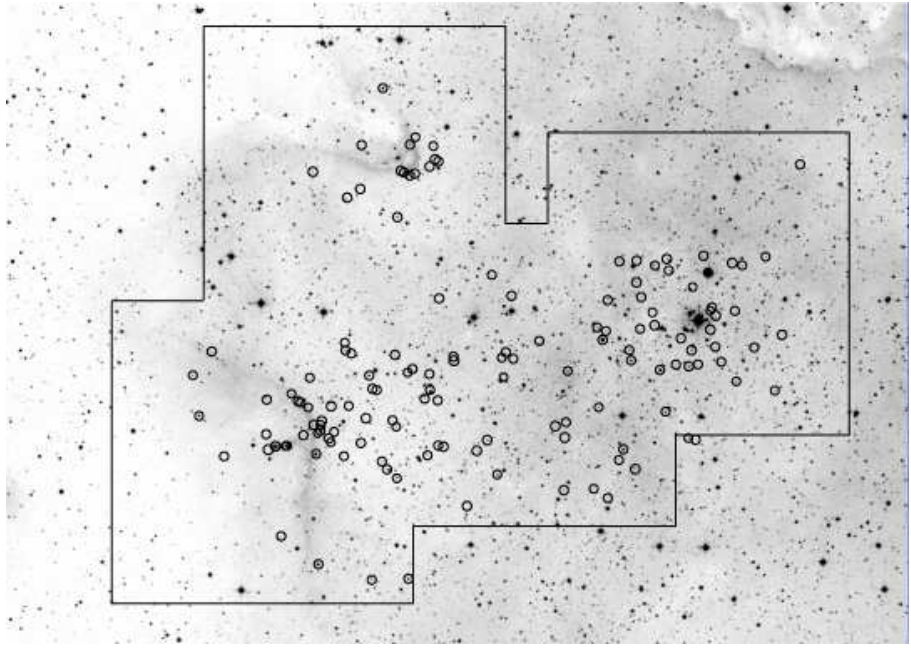

\begin{center}
\FigureFile(120mm,100mm){fig1.eps}
\end{center}
\caption{Spatial distribution of the detected H$\alpha$ emission stars overlaid on the DSS2 R plate.
 The observed field is indicated  by solid line.}
\label{fig1}
\end{figure}

\begin{figure}
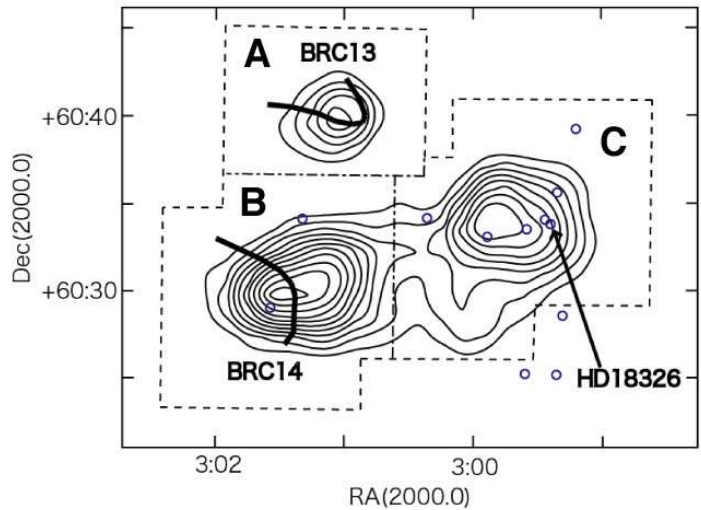

\begin{center}
\FigureFile(100mm,100mm){fig2.eps}
\end{center}
\caption{Isodensity contours of H$\alpha$ emission stars. 
OB stars in the observed field are denoted   by open circles.  
The isolines are drawn at intervals of 5$\times$ 10$^2$ stars per deg$^2$, starting from
10$^3$ stars per deg$^2$.
The boundary of the observed field is shown by dashed line. 
The prominent bright rims of BRC 13 and BRC 14 are sketched and shown by solid lines.
The vertical line separating the group B and C is defined at RA=3:00:40}
\label{fig2}
\end{figure}

Three groups of H$\alpha$ stars, A, B, and C, are separated about 8 pc each other, and the size of each 
group is 2.6 pc, 4.0 pc, and 3.8 pc for A, B, and C, respectively,  in FWHM.
The number of H$\alpha$ stars with small equivalent width ( $<$ 10 \AA) is also shown in table 2.
We could not find firm evidence showing the difference of the relative number of  
weak-lined T Tauri star candidates in three groups.

The exciting star lies at the $\sim$2 pc from the center of group C.
OB stars may affect the circumstellar environment of YSOs.
Recently, \citet{bal07} investigated the spatial distribution of Class II sources and O-type stars
in NGC 2244. They found that the effect of massive stars to the circumstellar disk of
Class II sources is  limited within 0.5 pc from the massive stars.
The displacement of the O-star and the peak of the group C may attribute to such a configuration.

\subsection{photometry}

We have made g' and i'-band photometry of 89 H$\alpha$ stars.
Figure 3 shows the position of our H$\alpha$ stars in a (g'-i',i') color-magnitude diagram.
By using the effective wavelength of i' and g'-band of SDSS system \citep{smi02} and
the equation of the parameterized Rv-dependent extinction law \citep{cad89}, assuming Rv=3.1,
we computed the extinction vector, which is shown with an arrow in figure 3.

\begin{figure}
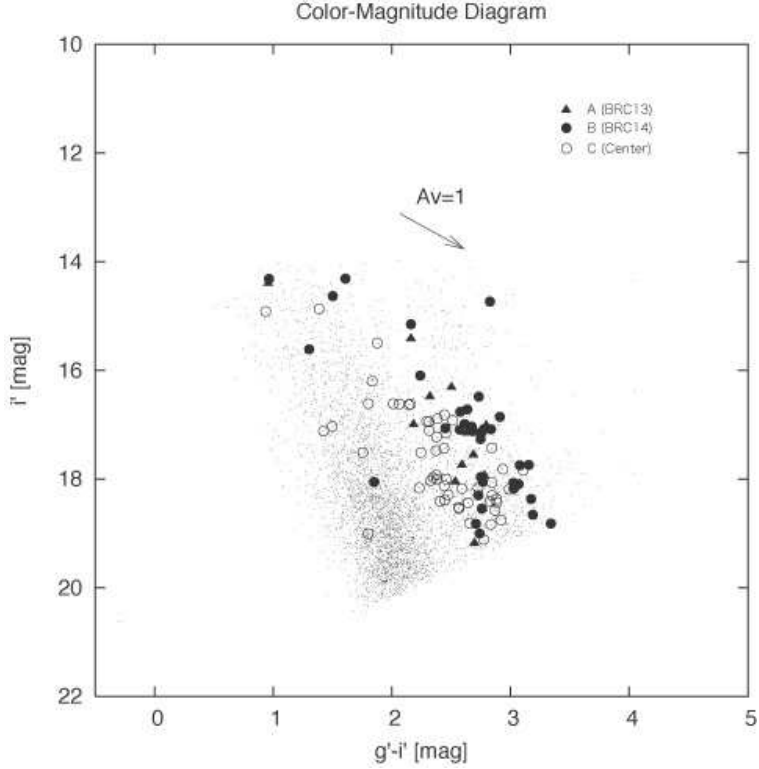

\begin{center}
\FigureFile(100mm,100mm){fig3.eps}
\end{center}
\caption{The color-magnitude diagram of all stars measured in the W5E region.
H$\alpha$ emission stars in the groups A,B, and C are marked as solid triangles, solid circles, and open
circles, respectively. }
\label{fig3}
\end{figure}

Although we  have no further spectroscopic information on intrinsic colors of H$\alpha$ stars, 
we used available JHK data to make NIR Two-Color Diagram (TCD) and  estimated the interstellar 
extinction of these stars.
Following the transformation equations by \citet{car01}, we converted NIR
colors of H$\alpha$ stars identified with the 2MASS Point Source Catalog into the CIT system.
H$\alpha$ stars  detected in J,H,K$_{s}$ bands with good quality (AAA) and the photometric error less than
0.1 magnitude are plotted in figure 4.
The H$\alpha$ emission stars in the groups A, B, and C are shown as solid triangles, solid circles, and open
circles, respectively.
We also plotted the main-sequence and giant (\cite{bes88}; thick solid line), and the 
dereddenned T Tauri (TTS) locus (\cite{mey97}; dashed line).
The reddening line (dotted line) is adopted from \citet{rie85}.
Tracing back H$\alpha$ stars from the position on the TCD to the  dereddened locus of 
TTS  along the reddening vector, we derived Av for each H$\alpha$ star.
Considering the error of the color, we have added several stars below 0.1 mag in J-H of TTS locus.
Their Av value are assumed to be zero. 
Six of seven stars at H-K$<$0.2 and J-H$<$0.5 are too faint to be  intermediate-mass stars (J$>$12),
indicating that most of them are background stars.

\begin{figure}
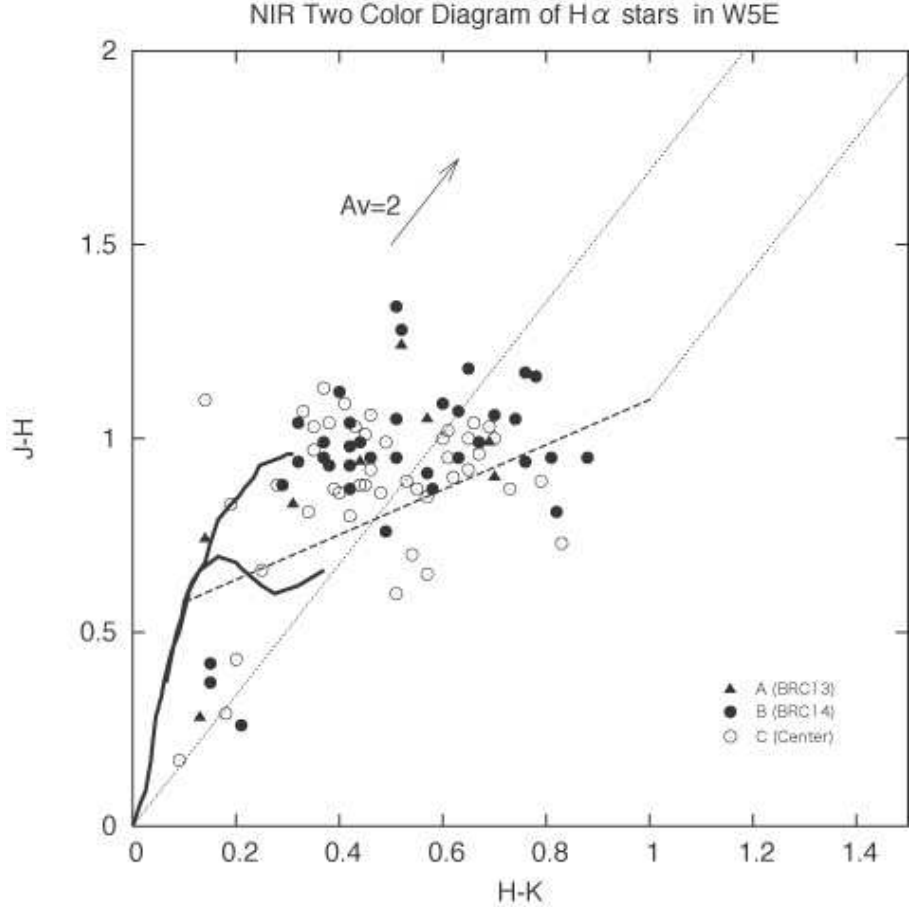

\begin{center}
\FigureFile(120mm,100mm){fig4.eps}
\end{center}
\caption{NIR two-color diagram of the H$\alpha$ emission stars in the 
W5E region.  H$\alpha$ stars in A,B, and C are shown as solid triangles, 
solid circles, and open circles, respectively.
}
\label{fig4}
\end{figure}

The mean Av for H$\alpha$ stars is 2.5 mag with a dispersion of 1.5 mag in
the combined groups A and B, while 1.9 mag with a dispersion of 1.3 mag in the group C.
\citet{han93} have studied the interstellar extinction properties of nine massive stars in Cas OB6.  
They have determined the total visual extinction from near-IR color excesses using a fit to 
the universal red to near-IR extinction curve. Av for Cas OB6 is between 1.9 and 3.1 mag, and
for IC 1848 (W5W) is around 2 mag. 
This is consistent with E(B-V)$=$0.69  for HD18326 \citep{sav85}, adopting Rv$=$3.1, which means 2.1 mag. 
As the foreground extinction estimated by \citet{han93} is about 2 mag for Cas OB6, there is 
almost no contribution to Av from the dust associated with the W5E region near the exciting source. 

Figure 5 is the extinction corrected color-magnitude diagram of the 53 H$\alpha$ stars.
The properties of our H$\alpha$ stars: serial number, identification number, position, photometric data, 
equivalent width of H$\alpha$ emission, Av, 2MASS name, and remarks are summarized in table 3.

\begin{figure}
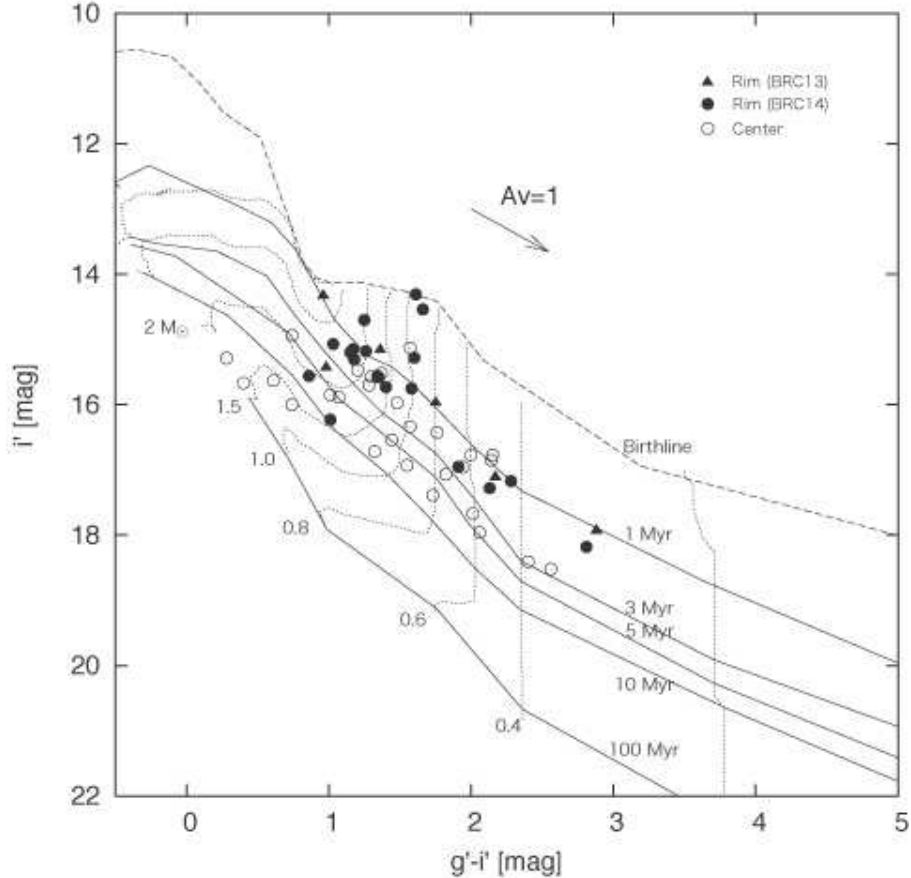

\begin{center}
\FigureFile(120mm,100mm){fig5.eps}
\end{center}
\caption{Extinction-corrected color-magnitude diagram of the H$\alpha$ emission stars in the 
W5E region.  
H$\alpha$ emission stars in the groups A,B, and C are shown as solid triangles, solid circles, and open
circles, respectively
the birthline and the 1, 3, 5, 10, 100 Myr isochrones of \citet{par99} are overlapped, as are the evolutionary
tracks for masses from 0.2 to 2.0 M$_\odot$, at an assumed distance of 2.3 kpc. The Av=1 mag reddening 
vector runs almost parallel to the isochrones.    }
\label{fig5}
\end{figure}

\section{Discussion}@

\subsection{Sequential formation of stars}

In order to estimate the mass and the age of each emission-line star,
we adopted the stellar models of \citet{par99} as pre-main sequence evolutionary tracks.
We converted their effective temperature and luminosity into the Johnson-Cousins system by using 
the conversion table by \citet{ken95} and then transformed them into the SDSS system by the empirical 
equation by \citet{jor06}.
The choice of theoretical pre-main sequence evolutionary tracks are sensitive to the
quantitative results.
Especially, the models have rather strong discrepancies in the regime of very low-mass stars. 
For example, \citet{luh03}  evaluate available evolutionary models by the observational constraints, 
e.g., coevality and mass of the young multiple systems, and confirmed the consistency.  
They extend the validity of theoretical models below a solar mass, and used the evolutionary model of \citet{par99}  
for M/M$_\odot$ $>$ 1 and the models of \citet{bar98} for M/M$_\odot$ $\leq$ 1.

We used the evolutionary models with metallicity Z=0.02 and no overshooting for the model of \citet{sie00} 
for comparison.
However, we found the  model of \citet{sie00}, which is converted to our photometric system by the table of 
\citet{ken95}, gives similar mass from the model of \citet{par99}. 
It gives somewhat older age in the mass range of 0.4--1.2 M$_\odot$ compared to the model of \citet{par99}, 
but not significant (20-30\%).  
Furthermore, the results are not sensitive to the extinction assumed because  the reddening 
vector lies nearly along the isochrones.

We derived masses and ages of H$\alpha$ stars from the color-magnitude diagram.
We ignored the effects of the unresolved binary companions and long term photometric 
variability \citep{bar05}  here.
Figure 6 shows the histogram of (a) masses and (b) ages of our H$\alpha$ stars for two regions, 
i.e., two bright-rimmed regions (group A and B) and the central region of HII region (group C).
The logarithmic mean mass of H$\alpha$ stars is 1.0 M$_\odot$ and 0.9 M$_\odot$ for the rim  
and central regions, respectively.  
The mass range of H$\alpha$ stars picked up in the present observations is between 0.2 M$_\odot$ and 2.0 M$_\odot$. 
The dispersion is rather large and there is no clear difference of mass between the regions.
The difference of age between the regions, in contrast,  is  clear in figure 6b. 
The logarithmic mean age of H$\alpha$ stars is 1 Myr and 4 Myr for the rim and central regions, respectively.
If we eliminate two stars nearly on the 100 Myr isochrone in figure 5 as the field stars, the above 
mean age for the central region changes to 3 Myr.
Nevertheless, the H$\alpha$ stars near the central exciting star are systematically older than 
those in the edge of HII region.

\begin{figure}
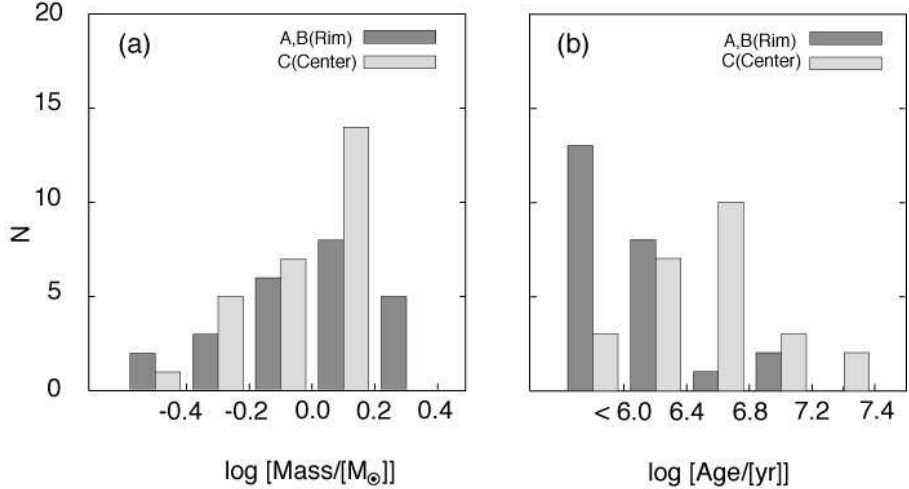

\begin{center}
\FigureFile(120mm,100mm){fig6.eps}
\end{center}
\caption{The distribution of (a) masses and (b) ages of H$\alpha$ emission stars, as read from the isochrones of figure 5.  }
\label{fig6}
\end{figure}

The youth of the sources in the rim region has been reported by many authors.
For example, \citet{kam03} showed the distribution of high-to-intermediate mass YSOs, and
found four luminous YSOs in our rim region.
The younger NIR  sources embedded in the BRC 14 have studied in detail 
by \citet{mat06}. They suggested the formation of the low-mass stars in the BRC 14 proceeds from outside
to the center of the cloud.
With an assumed velocity dispersion of 2 km s$^{-1}$, the compact aggregate could
have expanded to its current radius of $\sim$ 2 pc in 1 Myr.

Spitzer Space Telescope observations indicate that Class I sources are concentrated in the two embedded 
clusters near the edge of the molecular cloud
(BRC 13, BRC 14), suggesting that these are currently the most active and ongoing sites of star formation 
in the cloud \citep{all05}. 
On the other hand, Class II sources are widely distributed well off the molecular cloud into 
the adjacent HII region. 

The lower limit of the age of the HII region is estimated from the expansion age of the HII region, and 
the upper limit of the age is taken from the main sequence lifetime of the early type star,
giving between 1.7 and 3 Myr in age \citep{kam03}.
Thus, the stars in group A and B and younger sources in the bright-rimmed clouds are suggested to 
have been formed after the expansion of HII region.

For comparison, Class II sources, which have NIR colors explaining as the reddened TTS
between two parallel reddening line and above the TTS dereddened locus \citep{mey97} in figure 4,
are selected from the 2MASS catalog.  We again used the stars or sources with good quality, and 
photometric color error  $<$ 0.1 mag.
We have found 7, 27, and 31 NIR Class II sources in the group A, B, and C, respectively.
Figure 7 shows the spatial distribution of H$\alpha$ stars and NIR Class II sources overlaid on the $^{12}$CO (1-0) 
integrated intensity from the FCRAO Outer Galaxy Survey \citep{hey98}.
The overall distribution of NIR Class II sources is not evident as of the emission-line stars in figure 2, but is
in clear correspondence.
The sources in group B appear to be separated into two peaks, which correspond to the objects embedded in the bright rimmed
molecular cloud BRC 14 and adjacent to the west side of BRC 14.
It is naturally explained that the NIR observations selectively picked up the embedded sources in  BRC 14.
For example, \citet{mat06} estimated average of the extinction is 7--8 mag for the YSO candidates with 
NIR excess in BRC 14.

CO molecular gas mainly delineates the W5E HII region in the form of shell as seen in figure 7.
Furthermore,  another cloud component is the extending linear structure from BRC 14 to the 
neighboring small HII region S201, which has a bipolar structure at the 13' east of BRC 14.
There remains the possibility that stars in group C are in the background of the HII region.
However, their location probably reflects the original distribution of cloudy material
from which these stars formed.
We suggest that they are the fossil record of the pre-existing molecular cloud as
the $\lambda$ Orionis region extensively studied by \citet{dol02}.
The formation of stars proceeds sequentially from the center of HII region to the eastern bright rim.

\begin{figure}
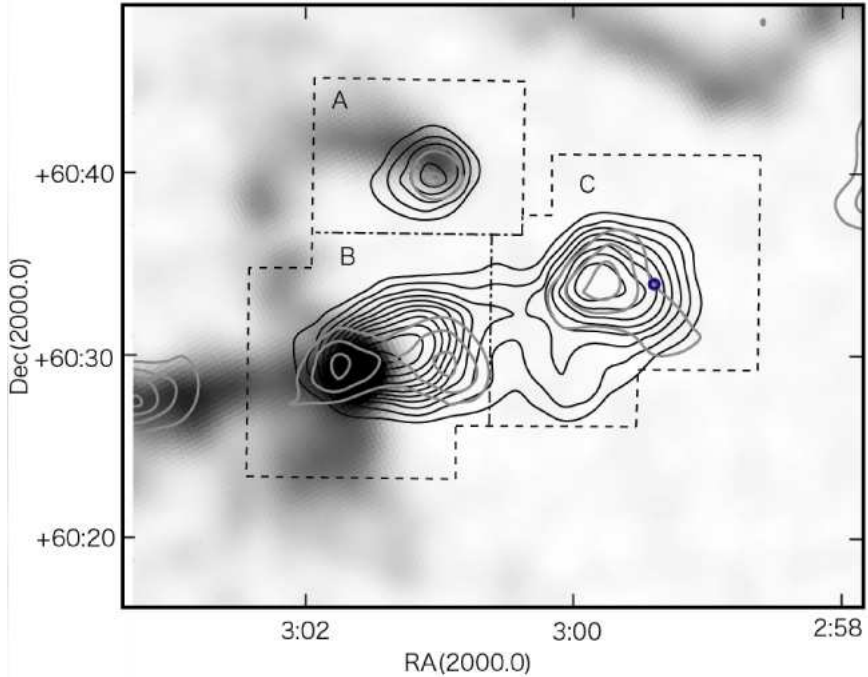

\begin{center}
\FigureFile(120mm,100mm){fig7.eps}
\end{center}
\caption{Isodensity contours of H$\alpha$ stars (solid lines) and NIR class II sources with NIR excesses (grey lines) 
overlaid on a grey-scale image 
of the $^{12}$CO (1-0)  integrated intensity between -57 and -32 km s$^{-1}$ from the FCRAO Outer Galaxy Survey. 
The isolines are drawn at intervals of 5$\times$ 10$^2$ stars per deg$^2$, starting from
10$^3$ stars per deg$^2$.
HD18328 is shown in open circle and the boundary of our observed field is shown by dashed line.}
\label{fig7}
\end{figure}

W5E contains one O star, HD18326, and 31 B stars within 30 arcmin from HD18326.
We have excluded one O star and four B stars associated with S201, reported from the NIR photometry by \citet{ojh04}.
Since the mass of the most massive star, HD18326, is 30 M$_\odot$ (O7V), the total mass of stars in W5E
is estimated to be 1300 M$_\odot$ \citep{lar03}.
If we adopt the multiple-part power law universal IMF by \citet{kro01}, 32 OB stars suggest that W5E contains 1300 stars 
with M $>$ 0.1 M$_\odot$, and
430 stars (33 \% ) should be  in the mass range of 0.2 $<$ M/M$_\odot$  $<$2.0 M$_\odot$,
which was covered by the present observations.
As the area that we have covered is approximately a third or a fourth of the whole W5E HII region, 
the average number of pre-main sequence stars expected from the Galactic-field universal IMF is 100--150 
in our observed area.
Although the present observations are not so deep and complete, we have picked up a considerable amount of 
low mass members in the HII region.

\subsection{Star formation in the central region of HII region}

\citet{math87}  suggested that OB stars in Cas OB6 are very young because
none of them evolved away from the main sequence.
Also the age of the W5E HII region was suggested to be  less than 3 Myr \citep{kam03}.
Considering the observational uncertainty and the age spread of star formation as a few Myr \citep{pre02},
the apparent age difference between the W5E HII region and H$\alpha$ stars in group C may not
deny the coeval formation of  OB stars in W5E and low-mass stars. 
Supposing that  H$\alpha$ stars near the exciting star and OB stars in W5E are the same generation, 
we could explain that they are  formed by the superbubble.
W4, the giant HII region with a cavity next to W5, is excited by the open cluster IC 1805 (OCL 352) 
which contains eight O-type stars \citep{mas95}.
IC 1805 is at the base of the Galactic chimney \citep{nor96}, and a loop of HII
reaching 1300 pc from Galactic midplane \citep{rey01}. 
\citet{oey05} suggested W3/W4 have a hierarchical triggering of star formation by 
superbubble.
The first generation corresponds to the superbubble of a 1300 pc loop and the Perseus chimney
of 6-20 Myr old.
Secondary, they triggered the formation of IC 1795 of 3--5 Myr age, and thirdly the OB stars in IC 1795  triggered 
the youngest (10$^4$ --10$^5$ yr) compact HII regions in W3.
\citet{car00}  speculated that the scattered and isolated low-mass clouds are the last remnants of a
giant molecular cloud that has been dispersed by OB stars.
Thus, we speculate the superbubble also triggered the formation of OB stars in W5E and dispersed TTS. 

While it is evident that  H$\alpha$ stars in the rim regions are younger than those near the exciting star (group C), 
the age difference between the stars in group C (4 Myr) and the exciting star ($<$ 3 Myr) may not be so clear. 
However, we should note that 4 Myr is a logarithmic mean age of the aggregates, and a considerable number
of low mass stars in group C might have been formed by the time. 
Additionally, the main sequence life time of the earliest stars in the neighbouring W5W region is 2.4 Myr 
(O6e; \cite{kam03}).  
Therefore, it is reasonable to consider that the most massive star of the W5E region was formed 
after the birth of many of  low mass stars in a cluster.
\citet{nal07} indicated the primordial turbulence in the cluster-forming region is 
replaced with the protostellar outflow-driven turbulence in the gravitational collapse time t$_{g}$ ($\sim$ 0.6 Myr).
They demonstrated that motions generated by outflow keep the region in 
a quasi-equilibrium state through 3D MHD numerical simulation. 
In their standard model, low mass stars started to form at 0.4 t$_{g}$, and more than
one hundred stars have formed at the end of the simulation (2 t$_{g}$).
The majority of the formed stars are clustered around the massive gas clumps.
They speculate that massive cluster-forming clumps gradually evolve toward a highly
condensed pivotal state, the instant of singular core formation, through a slow rate of star formation.
The massive core automatically reaches to be a dynamic collapse phase near the bottom of the cluster
potential well and forms massive star(s) rapidly  \citep{lin06}.
Although it is difficult to discriminate the formation timing between young stars and OB stars in HII region,
their idea is still very attractive in the case of the W5E HII region. 
The low mass star formation started in the pre-existing molecular cloud more than 4 Myr ago and they went through the
outflow phase as their protostellar activity.
Subsequently, massive stars were formed in the massive molecular clumps in the highly condensed state, 
and the UV radiation from the most massive star HD 18328 made the W5E HII region.
Then the primordial molecular gas associated with stars in group C was dissipated, and  
the young embedded star aggregate is gravitationally unbound after the gas leaves.
Finally, the ionizing radiation from the O star drove the clouds at the peripheries of HII region
to implode to form stars.
Now the most active star formation is in progress in the BRCs at the edge of the W5E HII region.

Much larger survey covering the entire complex and spectroscopic followup of H$\alpha$ stars 
to idenify them more clearly as PMS stars is needed
to clarify the star formation history of the whole W5 region.

\bigskip

The authors are supported by NAOJ for the use of the UH 2.2-m telescope for the observations.
This research has made use of the NASA/IPAC Infrared Science Archive, which is operated by the 
Jet Propulsion Laboratory,
California Institute of Technology, under contact with the National Aeronautics and Space Administration, and
used the facilities of the Canadian Astronomy Data Centre operated by the National Research Council of Canada 
with the support of the Canadian Space Agency. The research presented in this paper has used data from the Canadian 
Galactic Plane Survey a Canadian project with international partners supported by the 
Natural Sciences and Engineering Research Council.
This work was supported in part by a Grant-in-Aid for Scientific Research (17039011) from the
Ministry of Education, Culture, Sports, Science and Technology.






\end{document}